\documentclass{article}

\usepackage[dvips]{graphicx}
\usepackage{amscd}
\usepackage{xspace}
\usepackage{amstext}
\usepackage{amsmath}
\usepackage{amssymb,amsfonts,amsthm}
\usepackage{amscd}
\usepackage{hyperref}
\usepackage{bigstrut}
\usepackage{psfrag}
\newcommand{\p}{\partial}

\newcommand{\dd}{{\rm d}}

\begin{document}

{\large \noindent \bf  RELATIVITY PRINCIPLES IN 1+1 DIMENSIONS AND
\vspace{0.3cm} DIFFERENTIAL AGING REVERSAL }\vspace{1.5cm}


\begin{flushleft}
{\bf E. Minguzzi}\\
{ \it  Department of Applied Mathematics,}  { \it Florence
 University}, \\ { \it  Via S. Marta 3, I-50139 Florence, Italy} \\ { \it and INFN,
Piazza dei Caprettari 70, I-00186 Roma, Italy}\\
{ \it ettore.minguzzi@unifi.it } \vspace{1.5cm}
\end{flushleft}

\date{}

\noindent  We study the behavior of clocks in 1+1 spacetime assuming
the relativity principle, the principle of constancy of the speed of
light  and the clock hypothesis.
%
%
These requirements are satisfied by a class of Finslerian theories
parametrized by a real coefficient $\beta$, special relativity
being recovered for $\beta=0$. The effect of differential aging is
studied for the different values of $\beta$.
%
%
Below the critical values $\vert \beta \vert =1/c$ the differential
aging has the usual direction - after a round trip the accelerated
observer returns younger than the twin at rest in the inertial frame
- while above the critical values the differential aging changes
sign. The non-relativistic case is treated by introducing a formal
analogy with thermodynamics. \\ 




\section{\large INTRODUCTION}
Since their discovery the Lorentz transformations have been derived
in many ways
\cite{weinstock64,lee75,levyleblond76,macdonald81,field97}. The goal
was essentially that of reducing the number of postulates making
them logically and physically independent. In particular, some
authors recognized the need of an isotropy assumption in the
derivation \cite{berzi69,mahajan76,lee76,sardelis82}. In 1+1
dimensions the group of rotations is replaced by the group of space
reflections, and hence the isotropy assumption is replaced by a
parity assumption.

In this work we consider a 1+1-dimensional theory that  does not
satisfy the parity symmetry assumption although it is compatible
with the usual principles of special relativity. Its existence is
due to the presence (in 1+1 dimensions) of the Lorentz group
representation $\Lambda \to e^{ \beta \theta}$ where $\theta$ is
the rapidity and $\beta$ is an arbitrary constant. In this way one
can consider $ e^{\beta \theta} \Lambda$ as the coordinate
transformation that relates different inertial frames.  The
ultimate task is to study the behavior of clocks in the
corresponding generalized theories showing an interesting
phenomenon  that may occur varying the value of $\beta$: the
differential aging reversal.

This class of theories, dependent on the parameter $\beta$ were
studied  by Bogoslovsky \cite{bogoslovsky98} who also generalized
them to the higher 3+1-dimensional case. He considered a spacetime
$M$ endowed with a Finsler metric of the form (${\bf e}^2=1, {\bf
e}=cnst.$, $1/c\le \beta <0$)
\begin{eqnarray} \label{tau}
c^2 \dd \tau^2&=& (c \dd t-{\boldsymbol{e}}\cdot \dd {\bf
x})^{-2\beta c} (c^2\dd t^2-\dd {\bf x}^2)^{1+\beta c}
\end{eqnarray}
i.e. with an expression homogeneous of second  degree in the
differentials. Here $\beta$ is a universal constant with  the
dimension of the inverse of a velocity. Bogoslovsky used the
dimensionless constant $r=-\beta c$, but we prefer to consider
$\beta$ as a universal constant since it will allow us to obtain a
finite non-relativistic limit. Finsler metrics are a simple
generalization of Riemannian metrics (quadratic forms in the
differentials) and were considered by Riemann himself in his
seminal works on differential geometry. Finsler geometry is now a
mathematically mature subject \cite{round59}.

Physically, Finsler geometry provides a natural tool for the
description of anisotropic spacetimes \cite{asanov85}. For
instance, the previous metric is not invariant under the full
Poincar\'e group, as only the Lorentz transformations that send
the null vector $n^{\mu}=(1, \boldsymbol{e})$,  into a vector
proportional to itself may leave the metric unchanged
\cite{budden97,bogoslovsky99,bogoslovsky06}. It is clear that the
invariance under the group of Lorentz transformation is lost
because the spatial $SO(3)$ symmetry is broken into a $O(1)$
symmetry about ${\bf e}$. Nevertheless, the Einstein addition of
velocity rule holds true because the generalized Lorentz
transformations are just ordinary Lorentz transformations times a
factor\cite{bogoslovsky98,bogoslovsky99,bogoslovsky06}. The
breaking of rotational invariance reduces to a breaking of the
parity symmetry in the 1+1 case. Hence in this particular case the
theory is invariant under the entire proper orthochronous Lorentz
group.

A first interesting consequence of the metric  is the fact that
the light cones defined by $\dd \tau^2=0$ coincide with those of
Minkowski spacetime. Thus it does not introduce any change in the
usual causal structure. One can still identify the massive
particles with the future directed timelike worldlines.

The symbol $\tau$ for the metric line element suggests  that it
represents the time of a particle along the particle worldline.
This interpretation is correct and will be proved below. We shall
call that time {\em proper time} although it should not be
confused with the proper time of special relativity as it has a
different dependence on the velocity of the particle.

In this work we consider the clock problem of special relativity
where an accelerated observer makes a round-trip with respect to
an inertial observer. Their clocks previously synchronized are
compared at the end of the trip. The differential aging is the
difference between the proper time of the clock at rest in the
inertial frame and the proper time of the accelerating clock at
the final meeting event. Special relativity predicts that the
differential aging has always a positive sign independently of the
arbitrary motion performed during the round-trip by the
non-inertial observer. A similar result will be proved below in a
1+1 theory for which $\beta \in \mathbb{R}$ and such that when
$\beta$ crosses the values $\beta=\pm 1/c$ the differential aging
changes sign.

Notice that while in special relativity the expression $\tau=\int
\sqrt{1-v^2} \dd t$ with $\sqrt{1-v^2} \le 1$  implies $ \tau\le
 t$ in a round trip, in the generalized theories for $\beta
\ne 0$  the integrand may be above or below 1 depending on time
$t$. A priori, it is not clear that the balance of the
contributions should be independent of the round trip followed.

Our aim is also that of obtaining an expression for the
differential aging in terms of the acceleration history of the
non-inertial observer.  Indeed, as we shall see, given that
relation we will be able to prove the sign of the differential
aging using Holder's inequality. An analogous problem  was studied
in the special relativistic 1+1-dimensional case ($\beta=0$) in
\cite{minguzzi04c}. The idea is that the proper acceleration of
the inertial observer can be measured with a comoving
accelerometer and hence it is unambiguous and observable. On the
contrary  expressions like $\tau=\int \sqrt{1-v^2} \dd t$ involve
the velocity which depends on the convention of distant
simultaneity adopted. The differential aging versus acceleration
formulas, being independent of conventions, give more insights on
the physical spacetime structure. Connected to them is  the
problem of autonomous spacetime navigation that will not be
discussed here.

The problem of obtaining the differential aging from the
acceleration may become quite involved in the 3+1 case. Indeed, it
is difficult to obtain the motion from the decomposition of the
proper acceleration with respect to a Fermi-Walker transported
tetrad. Even the analogous Riemannian problem has not yet been
solved, although the uniqueness of the solution of the Frenet-Serret
equations is known since a long time. Thus, for computational
reasons we shall consider only the 1+1-dimensional case or, which is
the same, motions in 3+1 dimensions contained in a timelike plane.\\

\section{\large PROPER TIME IN FINSLER SPACETIMES} \label{vfe}
In the 1+1 spacetime $M$  the Finslerian  line element can be
conveniently written
\begin{eqnarray} \label{tau2}
c^2 \dd \tau^2&=& (c \dd t-\dd x )^{-2\beta c}
(c^2\dd t^2-\dd x^2)^{1+\beta c} , \label{tau2a}\\
&=& (c \dd t+\dd x )^{2\beta c} (c^2\dd t^2-\dd x^2)^{1-\beta c} ,
\label{tau2b} \\
&=&(c \dd t+\dd x )^{1+\beta c}(c \dd t-\dd x )^{1-\beta c} .
\end{eqnarray}
If $\vert \beta \vert < 1/c $ the causal structure coincides with
that of Minkowski spacetime. If $\beta >1/c$ the chronological
structure coincides with that of Minkowski spacetime and null
particles of velocity $\dd x/\dd t=c$ do not exist. If $\beta
<-1/c$ the chronological structure coincides with that of
Minkowski spacetime and null particles of velocity $\dd x/\dd
t=-c$ do not exist. The case $\vert \beta \vert = 1/c $ will be
discussed below.
%

Let $K'$ be a inertial frame in uniform motion with respect to $K$
at a velocity $v$. We define the rapidity $\theta$ with
\begin{equation}
e^{\theta/c}=\sqrt{\frac{1+v/c}{1-v/c}} ,
\end{equation}
or $\tanh \theta/c=v/c$.
 In our convention
$\theta$ has the dimension of a velocity. In this way we obtain the
right limit $\theta=v$  if $c \to \infty$, which is particularly
convenient as the relative frame velocity $v$ is an additive
parameter in non-relativistic physics.

The coordinate transformations that leave the metric invariant are
\begin{eqnarray}
ct'\!\!\!&=&\!\!\! e^{\beta \theta} [\cosh (\theta/ c) \,c t-\sinh
(\theta/ c) x]=\left[\frac{1+v/c}{1-v/c}\right]^{\beta c/2}
\!\!\!\!\frac{ct -\frac{v}{c} x}{\sqrt{1-v^2/c^2}}  ,\\
x'\!\!\!&=& \!\!\! e^{\beta \theta} [-\sinh (\theta/ c) \,c
t+\cosh (\theta/ c) x ]=\left[\frac{1+v/c}{1-v/c}\right]^{\beta
c/2}\!\!\!\! \frac{x -{v} t }{\sqrt{1-v^2/c^2}} ,
\end{eqnarray}
with $\theta \in \mathbb{R}$ and form a proper orthocronous
Lorentz group. The group structure reveals that all the inertial
reference frames are equivalent and  the equation $\dd \tau^2=0
\Rightarrow c^2\dd t^2-\dd x^2=0$  implies that the speed of light
is $c$ in every frame. Thus the two principles of special
relativity hold
\begin{itemize}
\item {\bf First principle}. A frame moving
uniformly with respect to an inertial frame is also an inertial
frame.
\item {\bf Second principle}. The
velocity of light $c$ is invariant under inertial frame changes.
\end{itemize}
However, since we do not assume invariance under reflection $x \to
-x$, special relativity in 1+1 dimensions is not recovered.

Now, notice that in its own coordinates, the motion of $K'$ reads
$x'=0$, and from the invariance of the Bogoslovsky metric we obtain
$\dd t'=\dd \tau$. Now, consider the arbitrary motion of an observer
$O$ along a timelike curve $\gamma$. At any instant there is  an
inertial frame $K'$ which is instantaneously at rest with $O$. We
use the locality principle \cite{mashhoon90,mashhoon90b} for clocks
measurements, sometimes referred to as the {\em clock hypothesis}
\cite{rindler77}. It states that the infinitesimal time elapsed in
an accelerating frame equals the infinitesimal time elapsed in the
local inertial frame $K'$. We conclude that the natural metric
parametrization $\tau$ represents the proper or natural time of the
non-inertial observer. This fact justifies the symbol $\tau$ for the
line element.

If $\vert \beta \vert=1/c$ then using Eqs.
(\ref{tau2a})-(\ref{tau2b}) we find $c \dd \tau=c \dd t+ {\rm
sgn}(\beta) \dd x$ which means that in this case there is no
differential aging and the transport of clocks leads to the
definition of a global time function  $ t+ {\rm sgn}(\beta) x/c$.
This global simultaneity definition, in which the simultaneity
hyperplanes coincide with the null hyperplanes orthogonal to $n$,
has been studied from different perspectives in
\cite{leutwyler78,duval85,duval91,budden97} and makes sense in
higher dimensional spacetimes as well.

\subsection{The non-relativistic limit}

The non-relativistic, $c \to + \infty$, limit of Eqs.
(\ref{tau2a})-(\ref{tau2b}) is
\begin{equation} \label{er}
 \dd \tau^2= e^{2\beta \dd x/\dd t }  \dd t^2 ,
\end{equation}
which is still a (degenerate) Finsler metric which seems to have
not received attention in the literature. We recall that in the
same limit $\theta=v$. Thanks to our dimensional choices for the
quantities $\beta$ and $\theta$ we have obtained a meaningful
non-relativistic limit. The invariance group is made by
transformations of the form
\begin{eqnarray}
t'&=& e^{\beta v} t
  , \label{abs}\\
x'&=&  e^{\beta v} (x-v t) ,
\end{eqnarray}
and can be regarded as the $c \to + \infty$ of their relativistic
counterparts. The classical Galilei Physics is obtained if $\beta=0$
in which case $\dd \tau=\dd t$ and again there is a global absolute
time $t$ measured by any (moving) clock. However, if $\beta \ne 0$
the natural time depends indeed on the path $x(t)$ followed,
although Eq. (\ref{abs}) clearly shows that the concept of
simultaneity is independent of the inertial frame chosen (i.e. we
have {\em absolute simultaneity} but not {\em absolute time}). \\

\section{\large THE TIME DILATION-ACCELERATION FORMULA}

We have obtained that the proper time $\tau$ of the non-inertial
observer is given by
\begin{eqnarray} \label{se1}
\tau &=& \int_{\gamma} (1+v/c)^{(1+\beta c)/2} (1-v/c)^{(1-\beta
c)/2} \dd t \label{bl}
\end{eqnarray}
if $c<+\infty$ and by
\begin{equation} \label{se2}
\tau = \int_{\gamma} e^{\beta v} \dd t
\end{equation}
if $c \to +\infty$. Special relativity corresponds to the case
($\beta=0,c <+\infty$), classical Galilei physics to the case
($\beta=0,c\to +\infty$). The equation (\ref{se2}) can be obtained
from the relativistic limit of (\ref{se1}).

Equations (\ref{se1})-(\ref{se2}) are not invariant under
reflections $x(t) \to - x(t)$. Thus a clock behaves differently
depending on the direction of motion, at least if we are not in the
special relativistic or classical case where $\beta=0$.  If one adds
a parity symmetry principle to the relativity principles and the
clock hypothesis then the usual special relativity or classical
physics is obtained. If this hypothesis is removed, introducing, for
instance, a small but non vanishing $\beta$, then alternative
theories for the behavior of clocks are obtained.

In this section we  consider  two clocks. The clock $C_0$ is at
rest in the origin of an inertial frame $K$. Its proper time is
denoted by $t$ and coincides with  K's inertial time coordinate.
The clock $C_1$ of proper time $\tau$ intersects $C_0$'s worldline
at two events. For semplicity assume that the two clocks have been
synchronized at the first meeting point in such a way that
$t=\tau=0$, and let $\bar{\tau}$ be $C_1$'s proper time at the
second meeting point. We are going to determine the differential
aging $t(\bar{\tau})-\bar{\tau}$ starting from $C_1$'s
acceleration $a(\tau)$ during the journey. Table \ref{table}
summarizes the results that will be proved in this section.

\begin{center}
\begin{table}
    \begin{center}
        \begin{tabular}{|c|c||c|c|}
            \hline
            & & &\\               &   {\small {\bf Einsteinian relativity}} & & {\small {\bf Galilean relativity}} \\
             & $0<c<+\infty$ & &$c=+\infty$\\
             & {\small Relative simultaneity} & & {\small Absolute simultaneity} \bigstrut\\
            \hline\hline
              & {\small Special relativity,} &  & {\small Classical physics,} \\
                 $\beta=0$  & {\small Proper time,} & $\beta=0$ &  $t(\bar{\tau}) = \bar{\tau}$, \\
                   & $t(\bar{\tau}) > \bar{\tau}$ &  &  {\small Absolute time:} $t$ \bigstrut\\
            \hline
           $0<\!\vert\beta\vert\!\!<1/c$ & $t(\bar{\tau}) > \bar{\tau}$ &  & 
      \bigstrut\\\cline{1-2}
         & $t(\bar{\tau}) = \bar{\tau}$,  &  &   \bigstrut\\
      $\vert\beta\vert =1/c$  & {\small Absolute time:}  &  $\beta \ne 0$ & $t(\bar{\tau}) < \bar{\tau}$ \bigstrut\\
        &  $t+\textrm{sgn}(\beta)x/c$ &  & \bigstrut\\\cline{1-2}
         $1/c\!<\!\vert\beta\vert\!\!<\!\!+\infty$ & $t(\bar{\tau}) < \bar{\tau}$ &    &\bigstrut\\\cline{1-2}
\hline
                              \end{tabular}
    \end{center}
    \caption{The parameters $c$ and $\beta$ determine the physical theory in 1+1 dimensions. Given a round trip in an inertial frame, $t(\bar{\tau})$ represents
    the time elapsed for a clock at rest in the inertial frame while $\bar{\tau}$ is the time elapsed in the accelerating frame. The differential aging is $\Delta=t(\bar{\tau})-\bar{\tau}$. At $\vert \beta \vert=1/c$ the differential aging changes sign (differential aging
    reversal).}
    \label{table}
\end{table}
\end{center}

From the differential relation between $\dd t$ and $\dd \tau$ we
find
\begin{eqnarray}
t(\tau)&=& \int^{\tau}_{0}e^{-\beta\theta(\tau')}\cosh[\theta(\tau')/c] \, \dd \tau' ,\label{e1} \\
x(\tau)&=&  \int^{\tau}_{0} e^{-\beta\theta(\tau')} c
\sinh[\theta(\tau')/c] \,
  \, \dd \tau'  \label{e2}.
\end{eqnarray}

Consider an infinitesimal boost in which the rapidity change is
$\Delta \theta$. With respect to the local inertial frame $K'$
instantaneously at rest with $C_1$, $\Delta \theta = \Delta v$ up
to higher order terms and thus one recovers $a= \dd \theta / \dd
\tau$ as in ordinary special relativity,
\begin{equation} \label{acc}
\theta(\tau)=\int^{\tau}_{0} a(\tau') \dd
\tau'+c\tanh^{-1}(v(0)/c) ,
\end{equation}
where $v(0)$ is  $C_1$'s velocity with respect to $K$ at the first
intersection point. At time $\bar{\tau}$, since $C_1$ returns in
$K$'s origin, we have $x(\bar{\tau})=0$. Using this relation in the
formula for $x(\bar{\tau})$ we find
\begin{equation} \label{v}
e^{2 \tanh^{-1}(v(0)/c)}=\frac{\int^{\bar{\tau}}_{0} e^{-(
\beta+\frac{1}{c})\int^{\tau'}_{0} a(\tau'') \dd \tau''} \, \dd
\tau'}{\int^{\bar{\tau}}_{0} e^{-(
\beta-\frac{1}{c})\int^{\tau'}_{0} a(\tau'') \dd \tau''} \, \dd
\tau'} .
\end{equation}
Plugging this in the equation for $t(\bar{\tau})$ we arrive at the
time dilation-acceleration formula
\begin{eqnarray}\label{tda}
t(\bar{\tau})&=&[\int^{\bar{\tau}}_{0} e^{\frac{1}{c}(1- \beta
c)\int^{\tau'}_{0}
a(\tau'') \dd \tau''} \, \dd \tau']^{\frac{1+\beta c}{2}}  \nonumber \\
&&\times \, [ \int^{\bar{\tau}}_{0} e^{-\frac{1}{c}(1+\beta
c)\int^{\tau'}_{0} a(\tau'') \dd \tau''} \, \dd \tau'
]^{\frac{1-\beta c}{2}} ,
\end{eqnarray}
that will be central in what follows. It generalizes a formula
given in \cite{minguzzi04c} for the case $\beta=0$.
 In particular if $v(0)=0$,
whatever the choice of sign, we have
\begin{equation}
t(\bar{\tau})=\int^{\bar{\tau}}_{0} e^{-( \beta \pm
\frac{1}{c})\int^{\tau'}_{0} a(\tau'') \dd \tau''} \, \dd \tau' .
\end{equation}

\subsection{The relativistic case}
Let us consider the $c< +\infty$ case.  Introduce the function
\begin{equation}
f(\tau)=e^{\frac{1}{2c}(1- \beta^{2}c^{2})\int^{\tau}_{0} a(\tau')
\dd \tau'} .
\end{equation}
The time dilation-acceleration equation can be rewritten
\begin{equation} \label{pr}
t(\bar{\tau})=[\int^{\bar{\tau}}_{0} f^{\frac{2}{1+\beta c}} \,
\dd \tau']^{\frac{1+\beta c}{2}} \, \, [ \int^{\bar{\tau}}_{0}
(f^{-1})^{\frac{2}{1-\beta c}}\, \dd \tau' ]^{\frac{1-\beta
c}{2}}.
\end{equation}
We have to consider different cases,

\subsubsection{The case
$\vert\beta\vert<1/c$.} Consider Eq. (\ref{pr}). From Holder's
inequality with $1/p=\frac{1+\beta c}{2}$, $1/q=\frac{1-\beta
c}{2}$, $p,q \in (1,+\infty)$ and $\frac{1}{p}+\frac{1}{q}=1$, it
follows $\bar{\tau} \le t(\bar{\tau})$. If $f$ is not proportional
to $f^{-1}$ the inequality is strict, which means that it is
strict unless $a(\tau)=0$, that is unless $C_{1}$ and $C_{0}$
follow the same trajectory for $\tau \in [0,\bar{\tau}]$.

\subsubsection{The case $\vert\beta\vert=1/c$.}
This case has been considered in section \ref{vfe} where we showed
that there is no differential aging and the transport of clocks
defines an absolute time $t + \text{sgn}(\beta) x/c$ with null
simultaneity slices (lightlike simultaneity).

 This is in any
case a good simultaneity definition as timelike particles move
forward with respect to the absolute time. The absolute time $t +
\text{sgn}(\beta) x/c$ is a Lorentz invariant as  follows from the
general expression of the Lorentz transformation for $\beta\ne0$,
\begin{eqnarray}
t'+x'/c&=& \large(\frac{1+v/c}{1-v/c}\large)^{\frac{\beta c-1}{2}} (t+x/c) ,\\
t'-x'/c&=& \large(\frac{1+v/c}{1-v/c}\large)^{\frac{\beta c+1}{2}}
(t-x/c) .
\end{eqnarray}

\subsubsection{The case $\vert\beta\vert>1/c$.}
We consider only the case $\beta>1/c$, the case $\beta < -1/c$
being analogous. Let $a \in \mathbb{R}$ be such that $\frac{\beta
c-1}{2}  < a < \frac{1+\beta c}{2}$ and let
\begin{eqnarray}
p&=&\frac{1+\beta c}{2a} \ \in \ (1,+\infty) , \\
q&=&\frac{2a}{\beta c -1} \ \in \ (1,+\infty)  .\\
g(\tau)&=&f(\tau)^{1/a} .
\end{eqnarray}
We can rewrite Eq. (\ref{pr}) as
\begin{equation} \label{opt}
\frac{t(\bar{\tau})}{\bar{\tau}}=\left\{\frac{(\frac{1}{\bar{\tau}}\int^{\bar{\tau}}_{0}
g^{1/p} \, \dd
\tau')^{p}}{(\frac{1}{\bar{\tau}}\int^{\bar{\tau}}_{0} g^{q}\, \dd
\tau' )^{1/q}} \right\}^{a} .
\end{equation}
Since $p,q>1$ we can use the inequalities
\begin{eqnarray}
(\frac{1}{\bar{\tau}}\int^{\bar{\tau}}_{0} g^{1/p} \, \dd
\tau')^{p} &\le& \frac{1}{\bar{\tau}}\int^{\bar{\tau}}_{0} g \,
\dd \tau' ,
\\
(\frac{1}{\bar{\tau}}\int^{\bar{\tau}}_{0} g^{q}\, \dd \tau'
)^{1/q} &\ge& \frac{1}{\bar{\tau}}\int^{\bar{\tau}}_{0} g \, \dd
\tau' ,
\end{eqnarray}
which are particular cases of Holder's inequality. These
inequalities are strict unless $g$ is a constant in which case
$a(\tau)=0$ in $[0,\bar{\tau}]$. Plugging these in Eq. (\ref{opt})
we obtain $t(\bar{\tau}) < \bar{\tau}$ unless the acceleration
vanishes in the interval in which case $t(\bar{\tau}) =
\bar{\tau}$ and hence $C_1$ is at rest in $K$.

\subsection{The non-relativistic case}
In the limit $c \to +\infty$, whatever the value of $\beta\ne 0$, we
have $\vert \beta \vert > 1/c$, thus we can suspect  that in the
non-relativistic case $t(\bar{\tau})< \bar{\tau}$. However, this
argument does not prove that in the limit it can not be
$t(\bar{\tau})= \bar{\tau}$. In order to obtain this result we have
to take the $c \to +\infty$ limit of Eq. (\ref{tda}). As we shall
see a nice formal correspondence with thermodynamics arises.

 Let
$E(\tau)=\int^{\tau}_{0} a(\tau') \dd \tau'$. The non-relativistic
$c \to +\infty$ limit  of Eq. (\ref{tda}) is
\begin{equation} \label{noo}
t(\bar{\tau})=(\int^{\bar{\tau}}_{0} e^{- \beta E(\tau')}\, \dd
\tau')e^{\beta\{ \frac{\int^{\bar{\tau}}_{0} e^{- \beta E (\tau')}
E(\tau')\, \dd \tau'}{\int^{\bar{\tau}}_{0} e^{- \beta E(\tau')}\,
\dd \tau'}\} } .
\end{equation}
This equation could have been obtained in a way analogous to that
followed in  the relativistic case by writing (see (\ref{se2}))
\begin{eqnarray}
t(\bar{\tau})&=&\int_0^{\bar{\tau}} e^{-\beta v(\tau)} \dd \tau ,\\
0=x(\bar{\tau})&=&\int_0^{\bar{\tau}} v(\tau) e^{-\beta v(\tau)}
\dd \tau ,\\
v(\tau)&=&v(0)+\int_0^{{\tau}} a(\tau') \dd \tau'.
\end{eqnarray}
Indeed, from the last two equations we  get $v(0)$ that replaced
into the first one gives (\ref{noo}).
 Equation (\ref{noo}) shows that in
the classical case ($\beta=0$) there is no differential aging. We
prove here that in the non-relativistic ($c=+\infty$) case  if
$\beta \ne 0$ there is always differential aging in the unusual
direction, $t(\bar{\tau}) < \bar{\tau}$ unless $C_0$'s and
$C_{1}$'s trajectories coincide in $\tau \in [0,\bar{\tau}]$.

 It is convenient to introduce the ``partition
function''
\begin{equation}
Z(\beta,
\bar{\tau})=\frac{1}{\bar{\tau}}\int_{0}^{\bar{\tau}}e^{-\beta
E(\tau')}\, \dd \tau' ,
\end{equation}
the ``probability density",
\begin{equation}
P(\tau)= \frac{\frac{1}{\bar{\tau}}e^{-\beta E(\tau)}}{
Z(\beta,\bar{\tau})},
\end{equation}
the ``mean energy" function,
\begin{equation}
\bar{E}(\bar{\tau})=\int_{0}^{\bar{\tau}} P(\tau')E(\tau') \dd
\tau'=-\frac{\p \ln Z}{\p \beta},
\end{equation}
and the ``entropy'' function,
\begin{equation} \label{entropy}
S=\ln Z-\beta \frac{\p \ln Z}{\p \beta} .
\end{equation}
Then
\begin{equation}
t(\bar{\tau})=\bar{\tau} e^{S} \qquad \textrm{or} \qquad S=\ln
\frac{t(\bar{\tau})}{\bar{\tau}}.
\end{equation}
These names and the letters $E$, $Z$, $S$, follow from the striking
formal analogy of these formulas with thermodynamics. The letter
$\beta$  was chosen keeping in mind this analogy. We shall consider
this correspondence only as a technical tool that will allow us to
use well known results from thermodynamics. From Eq. (\ref{v}) we
see that in the non-relativistic limit
\begin{equation}
v(0)=-\bar{E}=\frac{\p \ln Z}{\p \beta} ,
\end{equation}
which gives a kinematical interpretation for the mean energy
function. If $\beta=0$ one expects to recover the classical case.
Indeed we find that
\begin{equation}
v(0)=-\frac{1}{\bar{\tau}}\int_{0}^{\bar{\tau}}\int^{\tau}_{0}
a(\tau') \dd \tau' \dd \tau ,
\end{equation}
 or $\int_{0}^{\bar{\tau}} v(\tau) \dd \tau=0$, where $v(\tau)=v(0)+\int^{\tau}_{0} a(\tau') \dd \tau'$ is the
non-relativistic limit of (\ref{acc}). That is, the initial velocity
is determined by the non-relativistic condition that
$x(\bar{T})=x(0)=0$.
 Now, we want to determine if differential aging occurs.

Consider the ``energy variance",
\begin{equation}
\overline{(\Delta E)^{2}}=
\overline{(E-\bar{E})^{2}}=\overline{(E)^{2}}-\bar{E}^{2}
=-\frac{\p \bar{E}}{\p \beta}=\frac{\p^{2} \ln Z}{\p \beta^{2}}.
\end{equation}
Since, by definition, the variance is positive  we have
\begin{equation}
\frac{\p^{2} \ln Z}{\p \beta^{2}} \ge 0, \qquad -\frac{\p
\bar{E}}{\p \beta} \ge 0 .
\end{equation}
We obtain for the ``entropy"
\begin{equation}
\frac{\p S}{\p \beta}= -\beta \frac{\p^{2} \ln Z}{\p \beta^{2}}.
\end{equation}
Since its derivative, if different from zero, is negative for
positive $\beta$ and positive for negative $\beta$,  $S$ has a
maximum at $\beta=0$ where $S=0$ since $t=\bar{T}$ at that value. We
conclude that if $\tilde \beta \ne 0$ then $S(\tilde \beta) < 0$
unless $(\p S/\p \beta)(\beta)=0$  for all $\beta \in [0, \tilde
\beta]$. In this last case, however, $\overline{(\Delta E)^{2}}=0$,
that is $E(\tau)=\bar{E}$ is a constant and therefore there is no
acceleration. This completes the proof. \\

\section{\large CONCLUSIONS}
We have studied the behavior of clocks in a 1+1 dimensional
spacetime endowed with a Finsler metric. The theory satisfies  (a)
the relativity principle, (b) the principle of constancy of the
speed of light, and (c) the locality principle. We have not imposed
a parity symmetry assumption or an equivalent isotropy assumption.
The set of physical theories depends on two parameters, (i) the
speed of light $c$ and (ii) the constant $\beta$ such that the
coordinate transformation between inertial frames takes the form
$x'^{\mu}=e^{\beta \theta}\Lambda^{\mu}_{\ \nu} x^{\nu}$ where
$\theta=c \tanh^{-1}(v/c)$ is the rapidity and $v$ is the relative
velocity of the frames. We have shown that for a given $\beta \ne 0$
the differential aging has a sign that does not depend on the
arbitrary motion of the accelerated frame. This result  has been
obtained by applying Holder's inequality to a generalized time
dilation-acceleration formula. Moreover, we have proved that the
sign changes crossing the critical values $\beta=\pm 1/c$ for which
the differential aging vanishes. At these values  the transport of
clocks defines a global time variable or {\em absolute time} $t +
\textrm{sgn}(\beta)x/c$. The non-relativistic limit has also been
considered, showing that for $\beta\ne 0$ a formal correspondence
with thermodynamics arises where $\beta$ plays the role of an
inverse temperature.

These findings have been summarized in table \ref{table} which shows
that 2-dimensional theories, although based on the same fundamental
principles as higher dimensional theories, are richer than SR and
exhibit, through the  variation of $\beta$, a differential aging
reversal.

The problem whether these  results can be generalized to the $3+1$
dimensional theory considered by Bogoslovsky  and others remains
open and seems worth studying. \\ \\

 \noindent {\bf Acknowledgments.} The author has been supported by INFN, grant $\textrm{n}^{\circ}$ 9503/02.\\

%

\end{document}